\documentclass[a4paper]{article}
\usepackage[utf8]{inputenc}
\usepackage{INTERSPEECH2020}
\usepackage{cite}
\usepackage{times}
\usepackage{latexsym}
\usepackage{color}
\usepackage{listings}
\usepackage{textcomp}
\usepackage{caption}
\usepackage{mathtools}
\usepackage{wrapfig}
\usepackage{multirow}
\usepackage[hyphens]{url}
\usepackage[hidelinks]{hyperref}
\usepackage{cleveref}
\usepackage{amsmath,graphicx}
\usepackage{bookmark}
\usepackage{adjustbox}
\usepackage{siunitx}
\usepackage{setspace}
\usepackage{caption}
\usepackage{enumitem}

\DeclareUnicodeCharacter{00A0}{}

\newcommand{\blank}{\left<\operatorname{b}\right>}

\newcommand{\smalldots}{\makebox[0.8em][c]{...}}
\renewcommand{\paragraph}[1]{\textbf{#1}\hskip\parindent}

\makeatletter

\renewcommand{\section}{\@startsection
  {section}%
  {1}%
  {}%
  {-0.8\baselineskip}%
  {0.3\baselineskip}%
  {}}%

\renewcommand{\subsection}{\@startsection
  {subsection}%
  {2}%
  {}%
  {-0.1\baselineskip}%
  {0.1\baselineskip}%
  {}}%

\renewcommand{\subsubsection}{\@startsection
  {subsubsection}%
  {3}%
  {}%
  {-0.1\baselineskip}%
  {0.1\baselineskip}%
  {}}%

\g@addto@macro\normalsize{%
  \setlength\abovedisplayskip{3pt plus 2pt minus 1pt}
  \setlength\belowdisplayskip{3pt plus 2pt minus 1pt}
  \setlength\abovedisplayshortskip{2pt plus 2pt minus 1pt}
  \setlength\belowdisplayshortskip{2pt plus 2pt minus 1pt}
}

\makeatother

\setlist{
        itemsep=0pt,
        parsep=1pt plus 1pt minus 1pt,
        topsep=1pt plus 1pt minus 1pt,
        partopsep=0pt}

\graphicspath{{figures/}}

\title{A New Training Pipeline for an Improved Neural Transducer}
\name{Albert Zeyer$^{1,2}$, André Merboldt$^{1}$, Ralf Schlüter$^{1,2}$, Hermann Ney$^{1,2}$}
\address{
  $^1$Human Language Technology and Pattern Recognition,
  Computer Science Department, \\
  RWTH Aachen University, 52062 Aachen, Germany, \\
  $^2$AppTek GmbH, 52062 Aachen, Germany}
\email{\{zeyer, schlueter, ney\}@cs.rwth-aachen.de, andre.merboldt@rwth-aachen.de}

\begin{document}

\maketitle
\begin{abstract}
The \emph{RNN transducer} is a promising
end-to-end model candidate.
We compare the original training criterion
with the full marginalization over all alignments,
to the commonly used maximum approximation,
which simplifies, improves and speeds up our training.
We also generalize from the original neural network model
and study more powerful models,
made possible due to the maximum approximation.
We further generalize the output label topology
to cover RNN-T, RNA and CTC.
We perform several studies among all these aspects,
including a study on the effect of external alignments.
We find that the transducer model
generalizes much better on longer sequences
than the attention model.
Our final transducer model
outperforms our attention model
on Switchboard 300h
by over 6\% relative WER.
\end{abstract}
\noindent\textbf{Index Terms}: RNN-T, RNA, CTC, max.~approx.

\section{Introduction \& Related work}

\emph{End-to-end models} in speech recognition
are models with a very simple decoding procedure,
and often a simple training pipeline.
Usually the model directly outputs characters, subwords or words.
One of the earlier end-to-end approaches was
\emph{connectionist temporal classification (CTC)}
\cite{graves2006connectionist
}.
Most prominent is the \emph{encoder-decoder-attention model}
which has shown very competitive performance
\cite{zeyer2018:asr-attention,park2019specaugment,%
zeyer2019:trafo-vs-lstm-asr,tuske2020swbatt}.
Once the streaming aspect becomes more relevant,
or having a monotonicity constraint on the (implicit or explicit) alignment,
the global attention model needs to be modified.
Several ad-hoc solutions exists with certain shortcomings
\cite{chiu2017mocha,%
merboldt2019:local-monotonic-att}.
The \emph{recurrent neural network transducer (RNN-T) model}
\cite{graves2012seqtransduction,graves2013speechrnnt}
(or just \emph{transducer model})
is an alternative model
where the outputs
can be produced in a time-synchronous way,
and thus it is implicitly monotonic.
Because of this property,
RNN-T has recently gained interest
\cite{prabhavalkar2017comparison,%
rao2018rnnt,he2019streaming,li2019rnnt,jain2019rnnt,yeh2019trafot,%
andrusenko2020rnntchime,li2020streamingasr,%
zhang2020trafotransducer,%
hu2020pretrainrnnt,ghodsi2020rnntstateless,%
han2020contextnet,saon2020rnnt}.
RNN-T can be seen as a strictly more powerful generalization of CTC.

Several variations of RNN-T exists,
such as the recurrent neural aligner (RNA) \cite{sak2017neuralaligner},
monotonic RNN-T \cite{tripathi2019monotonicrnnt}
or hybrid autoregressive transducer (HAT) \cite{variani2020hat}.
The explicit time-synchronous modeling
also makes the alignment of labels explicit,
and requires a blank or silence label.
The alignment becomes a latent variable.
Most existing work keeps the model simple enough
such that the marginalization over all possible alignments
can be calculated efficiently via the forward-backward algorithm
\cite{graves2012seqtransduction}.
In case of RNA, an approximation is introduced.

Initializing the encoder parameters from another model
(such as a CTC model)
has often been done \cite{graves2013speechrnnt,rao2018rnnt,wang2018rnntchinese,hu2020pretrainrnnt}.
Initializing some of the decoder parameters from a language model
is common as well
\cite{rao2018rnnt,dong2018extendingrna,hu2020pretrainrnnt}.
Using an external alignment 
has been studied in \cite{hu2020pretrainrnnt}.

Differences in time-synchronous models 
(such as hybrid hidden Markov model (HMM) - neural network (NN)
\cite{bourlard1994hybrid,zeyer17:lstm})
vs.~label synchronous models 
(such as encoder-decoder-attention and segmental RNN)
w.r.t.~the alignment behavior are studied in \cite{beck2018:alignment}.
Time-synchronous decoding is also possible on
joint CTC-attention models \cite{moritz2019ctcatt}.

\section{Model}

Let $x_1^{T'}$ be the input sequence,
which is encoded by a bidirectional LSTM \cite{hochreiter1997lstm}
with time downsampling via max-pooling \cite{zeyer2018:asr-attention}
and optional local windowed self-attention \cite{lin2017selfatt}
\[ h_1^T = \operatorname{Encoder}(x_1^{T'}) . \]
Let $y_1^N$ be the target sequence,
where $y_n \in \Sigma$, for some discrete target vocabulary $\Sigma$,
which are byte-pair encoded (BPE) labels
\cite{sennrich2015neuralbpe,zeyer2018:asr-attention}
in our work. 
We define a discriminative model 
\[ p(y_1^N \mid x_1^{T'}) =
\sum_{\alpha_1^U : (T, y_1^N)}
\prod_{u=1}^{U} p(\alpha_u \mid \alpha_1^{u-1}, h_1^T) , \]
where $\alpha_u \in \Sigma' := \{\blank\} \cup \Sigma$,
where $\blank$ is the \emph{blank label}.
The \emph{output label topology $\mathcal{T}$ over $\Sigma'$}
defines the mapping on $t$, and generates the sequence $y_1^N$.
More specifically, the topology $\mathcal{T}$
defines
$\Delta t_\mathcal{T} (\alpha) \ge 0$ such that
$t_{u+1} = t_u + \Delta t_\mathcal{T} (\alpha_u)$,
and $t_1 = 1$, $t_{U} = T$,
and
$\Delta n_\mathcal{T} (\alpha) \ge 0$ such that
$n_{u+1} = n_u + \Delta n_\mathcal{T} (\alpha_u)$,
and $n_1 = 1$, $n_{U} = N$.
We study multiple variants of the label topology for $\alpha$.
Emitting a $\blank$ label will always consume a time frame,
and $\blank$ will be removed from the final output.
We study three variants:
\begin{itemize}
  \item \emph{CTC topology} \cite{graves2006connectionist}:
  Label emits time frame, repeated label will be collapsed.
    In this case, $U = T$,
    and $\Delta t \equiv 1$, $t_u = u$,
    and $\Delta n(\alpha_u) = \mathbf{1}_{\alpha_u \ne \blank \wedge \alpha_u \ne \alpha_{u-1}}$.
  \item \emph{RNA topology} \cite{sak2017neuralaligner}
  or \emph{monotonic RNN-T} \cite{tripathi2019monotonicrnnt}:
  Label emits time frame, repeats will not be collapsed.
    $U = T$,
    and $\Delta t \equiv 1$, $t_u = u$,
    and $\Delta n(\alpha) = \mathbf{1}_{\alpha \ne \blank}$.
  \item \emph{RNN-T topology} \cite{graves2012seqtransduction}:
  Label does not emit time frame, repeats will not be collapsed.
    $U = N + T$,
    and $\Delta t (\alpha) = \mathbf{1}_{\alpha = \blank}$,
    and $\Delta n(\alpha) = \mathbf{1}_{\alpha \ne \blank}$.
\end{itemize}

\begin{figure}[t]
  \centering
  \includegraphics[width=\linewidth]{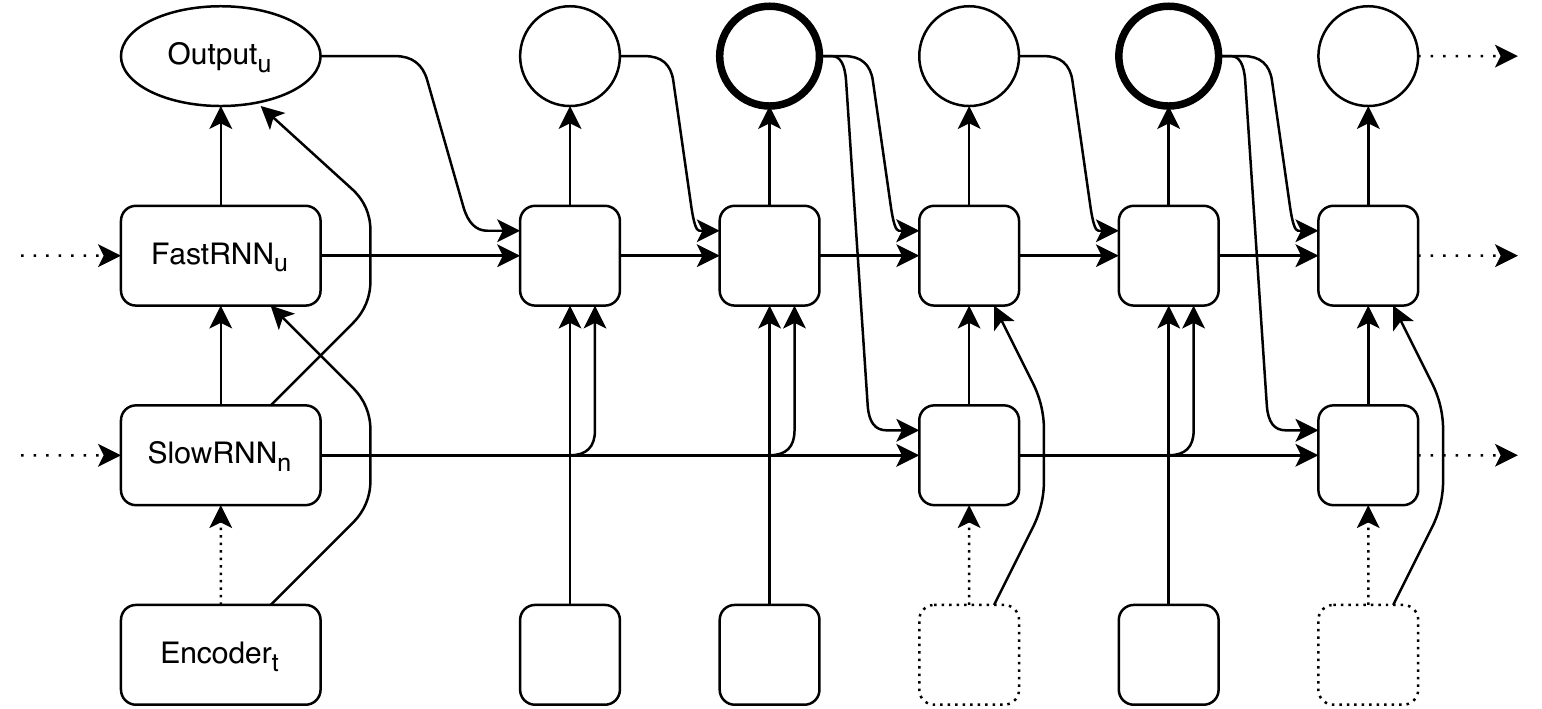}
  \caption{Unrolled decoder.
  The output labels are in $\{\blank\} \cup \Sigma$.
  A bold output depicts the emission of a new label in $\Sigma$
  (depending on the output label topology).
  SlowRNN is only updated when there is a new label.
  In case of RNN-T label topology,
  the encoder is only updated when there is no new label.
  In case of the original RNN-T model,
  SlowRNN has no dependency on the encoder.
  }
  \label{fig:unrolled-dec}
\end{figure}

We generalize from the common RNN-T and RNA model
and describe our decoder network in terms of a fast and slow RNN
\cite{schmidhuber1992chunker,liu2015multitime,%
chung2016hierarchicalrnn,mujika2017fastslowrnn,song2018dynframeskipping}.
The fast RNN iterates over $u \in \{1,\dots,U\}$,
while the slow RNN is calculated in certain sub frames
$n_u \in \{1,\dots,N\}$,
whenever there was a new generated $y$.
We visualize the unrolled decoder in \Cref{fig:unrolled-dec}.
The decoder for given frame $u$ is defined by
\begin{align*}
s^{\text{fast}}_u & := \operatorname{FastRNN}\!\left(
s^{\text{fast}}_{u-1},
s^{\text{slow}}_{n_u},
\alpha_{u-1},
h_{t_u}
\right) , \\
s^{\text{slow}}_{n_u} & := \operatorname{SlowRNN}\!\left(
s^{\text{slow}}_{n_u - 1},
\alpha_{u'-1},
h_{t_{u'}} \right), \\
u' & := \min \{k \mid k \le u, n_k = n_u \}   \tag{last emit} .
\end{align*}
Both the $\operatorname{SlowRNN}$ and the $\operatorname{FastRNN}$ are LSTMs in our baseline.
Note that we have $y_{n_u - 1} = \alpha_{u'-1}$,
and if we remove the dependency on $h$ in $\operatorname{SlowRNN}$,
and if we remove $\operatorname{FastRNN}$ and just set $s^{\text{fast}}_u = h_{t_u}$,
we get the original \mbox{RNN-T} model.
To get the probability distribution for $\alpha$ over $\Sigma'$,
we could use a single softmax,
as it was done for the original \mbox{RNN-T} and also RNA.
Our baseline splits $\Sigma$ and $\blank$ explicitly
into two separate probability distributions:
\begin{align*}
p(\alpha_u{=}\!\blank \mid \smalldots) & :=\sigma(\operatorname{Readout}^{b} (
s^{\text{fast}}_{u},
s^{\text{slow}}_{n_u}
)) , \\
p(\alpha_u {\ne}\! \blank \mid \smalldots) & \phantom{:}= \sigma(- \operatorname{Readout}^{b}
( s^{\text{fast}}_{u}, s^{\text{slow}}_{n_u} )) , \\
q(\alpha_u \mid \smalldots) & := \operatorname{softmax}_{\Sigma} ( \operatorname{Readout}^{y} (
s^{\text{fast}}_{u},
s^{\text{slow}}_{n_u}
)),
\ \alpha_u \in \Sigma \\
p(\alpha_u \mid \smalldots) & := p(\alpha_u {\ne}\! \blank \mid \smalldots) \cdot q(\alpha_u \mid \smalldots),
\ \alpha_u \in \Sigma
\end{align*}
where $\operatorname{Readout}$ is some feed-forward NN.
This can also be interpreted as a hierarchical shallow softmax.
HAT \cite{variani2020hat} uses a similar definition.
Note that the expressive power is equivalent to the single distribution:
\begin{align*}
q(\alpha_u \mid\smalldots) &
= \frac{p(\alpha_u \mid\smalldots)}{\sum_{\alpha'_u \in \Sigma} p(\alpha'_u \mid\smalldots) } ,
\ \alpha_u \in \Sigma \\
p(\alpha_u {\ne}\! \blank \mid \smalldots) & = 1 - p(\alpha_u {=}\! \blank \mid \smalldots) .
\end{align*}

\subsection{Training}

The model will be trained by minimizing the loss
\[ L := -\log p(y_1^N \mid x_1^T)
= -\log \sum_{\alpha_1^U : (T, y_1^N)} p(\alpha_1^U  \mid x_1^T) . \]
%
%
The sum $\sum_{\alpha_1^U}$ is usually solved via dynamic programming \cite{bellman57dynprog}
by iterating over $u \in \{1,\dots,U\}$.
When we have a dependence on individual $\alpha_u$ in the decoder,
it is not possible to calculate the sum $\sum_{\alpha_1^U}$ efficiently.
It is possible though to do an approximation in the recombination \cite{sak2017neuralaligner}.
The simplest approximation is to only use a single item of the sum,
specifically $\arg\max \alpha_1^U$, which is the well known maximum approximation,
which is done via the Viterbi alignment,
or some other external alignment,
as it is the standard approach for hybrid HMM-NN models
\cite{bourlard1994hybrid,zeyer17:lstm}.
We study two variants here:
\begin{itemize}
\item Exact calculation of the full sum (when possible).
\item Maximum approximation, together with fixed external alignment $\alpha_1^U$.
This is equivalent to frame-wise cross entropy training.
\end{itemize}


Using a fixed external alignment for the max.~approximation
has the disadvantage that we depend on a good external alignment,
which complicates the training pipeline.
But it comes with a number of advantages:
\begin{itemize}
\item We can train \emph{strictly more expressive models},
which can potentially be more powerful,
where the full sum cannot be calculated efficiently anymore.
\item The training itself is \emph{simpler}
and reduces to simple frame-wise cross entropy training.
This requires \emph{less computation} and should be faster.
\item It is \emph{more flexible},
and we can use methods like chunking \cite{zeyer17:lstm},
focal loss \cite{lin2017focalloss},
label smoothing \cite{szegedy2016labelsmth}.
\item In addition, maybe it is \emph{more stable}?
Or we get \emph{faster convergence} rate?
\end{itemize}

We also study the relevance of the type of external alignment.
This is a forced alignment on some other unrelated model
with the same output label topology.
This other model can be a weaker model,
only trained with the goal to generate the alignments.
We study several variants of models for this task.

%
%

\subsection{Decoding}

We use beam search decoding with a fixed small beam size
(12 hypotheses).
The hypotheses in the beam
are partially finished sequences $\alpha_1^u$ of the same length $u$.
The pruning is based on the scores $p(\alpha_1^u | x_1^T)$.
I.e.~the decoding is time-synchronous (in case $U = T$),
or synchronous over the axis $\{1,\dots,U\}$ \cite{saon2020rnnt}.
This is the same beam search algorithm and implementation
as for our attention-based encoder-decoder model \cite{zeyer2018:asr-attention}.
The only difference is that it runs over the axis $\{1,\dots,U\}$.

As an optimization of the beam search space,
we combine multiple hypotheses in the beam
when they correspond to the same partial word sequence (after BPE merging),
and we take the sum of their scores
(which is another approximation, based on the model).

\section{Experiments}

We use RETURNN \cite{zeyer2018:returnn} as the training framework,
which builds upon TensorFlow \cite{tensorflow2015}.
For the full-sum experiments on the RNN-T label topology,
we use warp-transducer%
\footnote{\tiny\url{https://github.com/HawkAaron/warp-transducer}}.
Our current full-sum implementation on the other label topologies
is a pure TensorFlow implementation
of the dynamic-programming forward computation.
Via auto-diff, this results in the usual forward-backward algorithm.
This is reasonably fast,
but still slower than a handcrafted pure CUDA implementation,
and also slower than the simple CE training.
Both training and decoding is done on GPU.
We present the training speeds in \Cref{tab:swb:training-speed}.
We publish all our code, configs and full training pipeline to reproduce our results%
\footnote{\tiny\url{https://github.com/rwth-i6/returnn-experiments/tree/master/2020-rnn-transducer}}.

All the individual studies are performed on the
Switchboard 300h English telephone speech corpus \cite{godfrey1992switchboard}.
We use SpecAugment \cite{park2019specaugment}
as a simple on-the-fly data augmentation.
We later compare to our attention-based encoder-decoder model \cite{zeyer2019:trafo-vs-lstm-asr}.

\begin{table}[t]
	\caption{On Switchboard 300h.
		For each model, label topology,
		loss (full-sum (FS) or frame-wise cross entropy (CE)),
		and loss implementation (pure TensorFlow (TF), or CUDA),
		we compare the \textbf{training time}
		on a single GTX 1080 Ti GPU.
		This measures the whole training, not just the loss calculation.
		CE training is without chunking.
	}
	\label{tab:swb:training-speed}
	\centering
	\setlength{\tabcolsep}{0.3em}
	\begin{tabular}{|l|c|c|c|c|c|}
		\hline
		\multirow{2}{*}{Model} & Label & \multirow{2}{*}{Loss} & Loss & \# params & time / epoch \\
		  & Topology    &       &  Impl.  & [M]           & [min]\\
		\hline
		\multirow{5}{*}{Transd.} & RNA & \multirow{4}{*}{FS} & \multirow{3}{*}{TF} & \multirow{5}{*}{147} & 306 \\ \cline{2-2}\cline{6-6}
		
		& CTC &  &  &  & 326 \\ \cline{2-2}\cline{6-6}

	& \multirow{2}{*}{RNN-T} &  &  &  & 333\\ \cline{4-4}\cline{6-6}

	&  &  & CUDA &  & 219\\ \cline{2-4}\cline{6-6}

& CTC & \multirow{2}{*}{CE} & \multirow{2}{*}{TF} &  & 160\\
\cline{1-2}\cline{5-6}

Attention & $-$ &  &  &  162  & 138 \\
		\hline
	\end{tabular}
\end{table}

\subsection{Full-sum vs.~frame-wise cross entropy}

We compare full-sum (FS) vs.~frame-wise cross entropy (CE)
in \Cref{tab:swb:fullsum-vs-approx}.
We observe that the full-sum training is more unstable,
esp.~in the beginning of the training,
and leads to worse performance
within the same amount of training time.
We do not count the time to get the external alignment,
so the comparison might not be completely fair.
Chunking also has a positive effect on the CE training,
as we will show later in \Cref{tab:swb:ablation}.


\begin{table}[t]
	\caption{On Switchboard 300h, transducer model,
	without external LM.
		Comparison of \textbf{full-sum (FS)} training
		and \textbf{frame-wise cross entropy (CE)} training
		via a fixed external alignment.
		All models are trained for 25 epochs
		and share the same network topology,
		which has no label feedback to allow FS training.
		CE training uses CTC alignments,
		where label repetition is enabled for CTC-Vit and disabled for RNA-Vit.
		CE training uses chunking.
	}
	\label{tab:swb:fullsum-vs-approx}
	\centering
	\setlength{\tabcolsep}{0.3em}
	\begin{tabular}{|c|c|c|c|c|c|}
		\hline
		Label & Training & \multicolumn{4}{c|}{WER[\%]}\\
		Topology & Criterion  & \multicolumn{3}{c|}{Hub5'00} & Hub5'01\\
		&  &  SWB & CH & $\Sigma$ & $\Sigma$\\
		\hline
		\multirow{2}{*}{RNA} & FS  & 11.5 & 23.4 & 17.5 & 16.5 \\ \cline{2-6}

& CE & 10.1 & 20.4 & 15.2 & 14.8\\
		\hline

		\multirow{2}{*}{CTC} & FS  & 15.0 & 24.6 & 19.8 & 20.1\\ \cline{2-6}

		& CE & 10.5 & 20.6 & 15.6 & 15.3\\
		\hline

		RNN-T & FS   & 11.6 & 22.3 & 17.0 & 16.4\\
		\hline
	\end{tabular}
\end{table}

We study the influence of the external alignment for the frame-wise CE training
in \Cref{tab:swb:alignments}.
We see that a standard CTC model can be used to generate an alignment,
but we also see that other models produce better alignments for our purpose.
Specifically, using a transducer model (trained from scratch with full-sum)
to generate the alignment seems to work best.
This is as expected, as this is the most consistent setup.

\begin{table}[t]
  \caption{On Switchboard 300h, WER on Hub 5'00.
  CE-trained transducer B1 and B2 models
  (\Cref{sec:exp:variations}),
  always with CTC label topology,
  without external LM,
  with randomly initialized parameters,
  trained for 25 epochs.
  Comparing \textbf{alignments}
  (specifically the models used to get the alignments).
  The alignment model
  was also always trained for 25 epochs.
  }
  \label{tab:swb:alignments}
  \centering
  \setlength{\tabcolsep}{0.3em}
  \begin{tabular}{|l|c|c|c|c|c|c|c|c|}
    \hline
    Alignment model & \multicolumn{2}{c|}{WER[\%]} \\ \cline{2-3}
    & B1 & B2 \\
    \hline
CTC-align 4l & 14.7 & 14.3 \\ \hline
CTC-align 6l & 14.7 & 14.5 \\ \hline
CTC-align 6l with prior (non-peaky) & 15.4 & 14.9 \\ \hline
CTC-align 6l, less training & 14.6 & 14.6 \\ \hline

Att.-based enc.-dec. + CTC-align & 14.4 & 14.2 \\ \hline


Transducer-align & 14.2 & 14.1 \\
\hline
\end{tabular}
\end{table}

%

%


\subsection{Ablations and variations}
\label{sec:exp:variations}


Along our research on training transducer models,
we came up with many variants,
until we eventually ended up with the baselines B1 and B2.
Both transducer baselines
use
the CTC label topology
with a separate sigmoid for the blank label
(similar as in \cite{variani2020hat}).
We use CE training using the fixed alignment CTC-align 6l
(as in \Cref{tab:swb:alignments}).
Sequences of the training data are cut into chunks \cite{zeyer17:lstm}.
We use
focal loss \cite{lin2017focalloss},
an additional auxiliary CTC loss on the encoder
(for regularization \cite{hori2017attctc,zeyer2018:asr-attention}),
dropout \cite{srivastava2014dropout},
dropconnect \cite{wan13dropconnect} (weight dropout)
for the FastRNN,
and switchout \cite{wang2018switchout}
(randomly switch labels for label feedback).
B1 uses local windowed self-attention,
while B2 does not, which is the only difference between B1 and B2.
Some of these tricks were copied from our hybrid HMM-NN model \cite{zeyer17:lstm}.
Based on these baselines,
we want to see the effect of individual aspects
of the model or training.
We summarized the variations and ablations in
\Cref{tab:swb:ablation}.
We see that chunked training greatly helped
for CE training, which is consistent with the literature
\cite{zeyer2017:ctc}.
We find that the results about label feedback are not conclusive.
Having the separate sigmoid for blank seems to help.


\begin{table}[t]
  \caption{On Switchboard 300h, WER on Hub5'00.
  \textbf{Ablations and variations}.
  Using transducer baselines B1 and B2
    (see \Cref{sec:exp:variations} for details),
  without external LM.
  B2 is exactly the B1 baseline without local windowed self-attention.
  }
  \label{tab:swb:ablation}
  \centering
  \setlength{\tabcolsep}{0.3em}
  \begin{tabular}{|l|c|c|c|c|c|c|c|c|}
    \hline
    Variant & \multicolumn{2}{c|}{WER[\%]} \\ \cline{2-3}
    & B1 & B2 \\
    \hline
    Baseline & 14.7 & 14.5 \\ \hline
  No chunked training & 16.3 & 15.7 \\ \hline
    No switchout & 15.0 & 14.5 \\ \hline
    SlowRNN always updated (not slow) & 14.8 & 14.8 \\ \hline
  No SlowRNN & 14.8 & 14.7 \\ \hline
    No attention & 14.5 & * \\ \hline
   FastRNN dim 128 $\rightarrow$ 512 & 14.3 & 14.5 \\ \hline
No encoder feedback to SlowRNN & 14.9 & 14.7 \\ \hline
 + No FastRNN label feedback (like RNN-T) & 14.9 & 14.5 \\ \hline
+ No FastRNN (exactly RNN-T) & 15.2 & 15.1 \\ \hline
    No separate blank sigmoid & 14.9 & 14.9 \\
    \hline
  \end{tabular}
\end{table}

\subsection{Output label topology}

We compare the output label topology in \Cref{tab:swb:fullsum-vs-approx}.
We find that the RNN-T topology seems to perform best,
followed by RNA,
and CTC is worse.
We note that this result is inconsistent to earlier results,
where CTC looked better than RNA.
However, when we repeat the RNA vs.~CTC comparison
on the B2 model with CE training,
we also see that RNA performs better (14.2\% vs 14.5\% WER on Hub5'00).
For simplicity, we did not follow the RNN-T topology further in this work.
Also, because of our earlier results, we focused more on the CTC topology.

\subsection{Importing existing parameters}

For faster and easier convergence,
it can be helpful to import existing model parameters
into our RNA model.
If we do not use the full-sum training,
we anyway make use of an external alignment,
which comes from some other model,
so it might make sense to reuse these parameters.
We collect our results in \Cref{tab:swb:enc-import}.
We see that the CTC model parameters
(of the same model which was used to create the alignment)
seem to be suboptimal,
and training from scratch performs better.
The encoder of an attention-based encoder-decoder model
seems to be very helpful.
Importing the model itself,
i.e.~effectively training twice as long, helps just as much.
However, we can also use the attention-based encoder-decoder model
with an additional CTC layer on-top of the encoder
to generate the alignments
as shown in \Cref{tab:swb:alignments}.
We also tried to initialize the SlowRNN with the parameters of a LM
but this had no effect.

\begin{table}[t]
  \caption{On Switchboard 300h, WER on Hub5'00.
  Varying the \textbf{imported model params},
  for transducer baseline models B1 and B2,
  with CTC topology,
  trained with CE using a fixed external alignment
  (CTC-align 6l),
  without external LM.
  Trained  for 25 epochs,
  with randomly initialized parameters,
  except of the imported ones.
  The imported models themselves are also trained for 25 epochs.
  }
  \label{tab:swb:enc-import}
  \centering
  \setlength{\tabcolsep}{0.3em}
  \begin{tabular}{|l|c|c|c|c|c|c|c|c|}
    \hline
    Imported model params & \multicolumn{2}{c|}{WER[\%]} \\ \cline{2-3}
    & B1 & B2 \\
    \hline
None & 14.7 & 14.5 \\ \hline
CTC as encoder & 15.4 & 15.5 \\ \hline
Att. encoder & 14.2 & 13.9 \\ \hline
Transducer (itself) & 13.7 & 13.6 \\
\hline
\end{tabular}
\end{table}

\subsection{Beam search decoding}

We study different beam sizes, and compare the attention model
and our RNA model.
For RNA, we also implemented a variation of the beam search
where hypotheses corresponding to the same word sequence
(i.e.~after collapsing label repetitions, removing blank, and BPE merging)
were recombined together by taking their sum (in log space)
and only the best hypothesis survives.
The results are in \Cref{tab:swb:beam-sizes}.
In all cases, the WER seems to saturate for beam size $\ge 8$.

\begin{table}[t]
	\caption{On Switchboard 300h, WER on RT03S.
	The transducer uses the CTC-label topology.
	  Without external LM.
		Comparison of performance on \textbf{different beam sizes}.
		We optionally recombine hypotheses in the beam corresponding
		to the same word sequence
		(after collapsing repetitions, removing blank, and BPE merging).
		}
	\label{tab:swb:beam-sizes}
	\centering
	\setlength{\tabcolsep}{0.2em}
	\begin{tabular}{|c|c|c|c|c|c|c|c|c|c|}
		\hline
		Model & Merge & \multicolumn{8}{c|}{WER[\%]} \\ \cline{3-10}
		&& \multicolumn{8}{c|}{Beam size} \\
		&& 1 & 2 & 4 & 8 & 12 & 24 & 32 & 64 \\
		\hline
		Att. & \multirow{2}{*}{no} & 17.9 & 17.0 & 16.7 & 16.6 & 16.6 & 16.5 & 16.6 & 16.5  \\ \cline{1-1}\cline{3-10}
		\multirow{2}{*}{Transd.} &  & 16.8 & 16.4 & 16.2 & 16.2 & 16.2 & 16.2 & 16.2 & 16.2  \\ \cline{2-10}
		 & yes & 16.8 & 16.3 & 16.0 & 15.9 & 15.9 & 15.9 & 15.9 & 16.0  \\ \hline
	\end{tabular}
\end{table}

\subsection{Generalization on longer sequences}

It is known that global attention models do not generalize well
to longer sequences than seen during training
\cite{rosendahl2019:pos_enc,narayanan2019longform,chiu2019longform}.
Esp.~the attention process has problems with this.
The alignment process in the transducer is explicit,
and this aspect should have no problems in generalizing to any sequence length.
To analyze, during recognition,
we concatenate every $C$ consecutive seqs.~within a recording
and thus increase the avg.~seq.~lengths.
We show the results in \Cref{tab:swb:longer-seqs}.
We report the WER on RT03S to minimize overfitting effects.
Both models
degrade with longer sequences,
but the attention model performs much worse,
and the transducer model has generalized much better.
The small degradation of the transducer could also be explained
by unusual sentence boundaries.
We note that this small degradation looks better than reported previously (relatively)
\cite{narayanan2019longform,chiu2019longform,chiu2020rnntgeneralization}.

\begin{table}[t]
	\caption{On Switchboard 300h, WER on RT03S.
	The transducer uses the CTC output topology.
  Without external LM, beam size 12.
		Comparison of performance on \textbf{varying sequence lengths},
		by concatenating every $C$ consecutive seqs.~(only in recog.).
		}
	\label{tab:swb:longer-seqs}
	\centering
	\setlength{\tabcolsep}{0.3em}
	\begin{tabular}{|S[table-format=3]|c|c|c|c|c|c|c|}
		\hline
		$C$ & \multicolumn{2}{c|}{Seq.~length [secs]} && \multicolumn{2}{c|}{WER[\%]}  \\ \cline{5-6}
		& mean$\pm$std & min-max && \multicolumn{1}{c|}{Att.} & \multicolumn{1}{c|}{Transd.} \\
		\hline
		1 & \phantom{00}2.71$\pm$\phantom{0}2.38 & 0.17-\phantom{0}34.59 && 16.5 &  16.0 \\ \hline
		2 & \phantom{00}7.83$\pm$\phantom{0}4.96 & 0.42-\phantom{0}63.70 && 16.8 &  16.1 \\ \hline
		4 & \phantom{0}17.74$\pm$\phantom{0}9.25 & 0.42-\phantom{0}91.62 && 17.9 &  16.3 \\ \hline
		10 & \phantom{0}45.57$\pm$19.93 & 0.74-126.52 &&29.3 & 16.7 \\ \hline
		20 & \phantom{0}86.37$\pm$39.58 & 0.74-194.10 && 51.9 & 17.1 \\ \hline
		30 & 122.10$\pm$57.28 & 1.17-297.15 && 65.1 & 18.1 \\ \hline
		100 & 290.58$\pm$35.50 & 8.54-309.14 && 94.8 & 18.2 \\
		\hline
	\end{tabular}
\end{table}

\subsection{Overall performance}

Our final transducer model is based on B1 but with RNA label topology
and better transducer-based alignment.
We compare to our attention model
and to other results from the literature
in \Cref{tab:swb:overall}.
Our final transducer model
performs better than our attention model,
although it needs the preprocessing step
to get an alignment.
We observe that many other works train for much longer,
and there seems to be a correlation between training time and WER.

\begin{table}[t]
  \caption{On Switchboard 300h,
  comparing final results of our \textbf{transducer model}
  with RNA label topology
  to our \textbf{attention model},
  and to other attention models from the \textbf{literature}.
  One big difference in varying results
  is the different amount of training time,
  which we state as number of epochs.
  }
  \label{tab:swb:overall}
\begin{adjustbox}{width=1.\width,center}
\setlength{\tabcolsep}{0.17em}
  \begin{tabular}{|c|c|c|c|c|S[table-format=2.1]|S[table-format=2.1]|S[table-format=2.1]|S[table-format=2.1]|S[table-format=2.1]|c|c|}
    \hline
    Work & \multicolumn{2}{c|}{Label} & \#Ep & LM & \multicolumn{5}{c|}{WER[\%]} \\
    &  Type & \# && & \multicolumn{3}{c|}{Hub5$^{00}\!$} & \multicolumn{1}{c|}{Hub5$^{01}\!$} & RT$^{03}\!$ \\
    && && & SWB & CH & $\Sigma$ & $\Sigma$ & $\Sigma$ \\
    \hline
    \cite{zeineldeen:icassp20} & Phone & 4.5k & \phantom{0}13 & yes & 9.6 & 18.5 & 14.0 & 14.1 & \\
    \hline\hline

    \cite{zeyer2019:trafo-vs-lstm-asr} & \multirow{3}{*}{BPE} & 1k & \phantom{0}33 & \multirow{2}{*}{no} & 10.1 & 20.6 & 15.4 & 14.7 & \\
    \cline{1-1}\cline{3-4}\cline{6-10}

    \cite{nguyen2019improving} &  & 4k & \phantom{0}50 &  & 8.8 & 17.2 & 13.0 && \\
    \cline{1-1}\cline{3-10}

    \cite{karita2019trafo} &  & 2k & 100 & \multirow{2}{*}{yes} & 9.0 & 18.1 & 13.6 & & \\
    \cline{1-4}\cline{6-10}

    \cite{wang2020phonebpe} & BPE$^{\text{\tiny Ph}}$ & 500 & 150 &  & 7.9 & 16.1 & && 14.5 \\
    \cline{1-10}

    \cite{tuske2020swbatt} & BPE & 600 & 250 & \multirow{2}{*}{no} & 7.6 & 14.6 &&& \\
    \cline{1-4}\cline{6-10}

    \cite{park2019specaugment} & WPM &1k & 760 &  & 7.2 & 14.6 & && \\
    \hline
    \hline
    \multicolumn{1}{|c}{Ours} & \multicolumn{9}{c|}{} \\ \hline
    \multirow{2}{*}{Att.} & \multirow{4}{*}{BPE} & \multirow{4}{*}{1k} &  \phantom{0}25 & \multirow{4}{*}{no} & 9.2 & 21.1 & 15.2 & 14.2 & 17.6  \\
    \cline{4-4}\cline{6-10}

    & &&  \phantom{0}50 & & 8.7 & 19.3 & 14.0 & 13.3 & 16.6 \\
    \cline{1-1}\cline{4-4}\cline{6-10}

\multirow{2}{*}{Transd.}    & && \phantom{0}25 & & 9.4 & 18.7 & 14.1 & 14.1 & 16.7\\
    \cline{4-4}\cline{6-10}

    & && \phantom{0}50 & & 8.7 & 18.3 & 13.5 & 13.3 & 15.6 \\
    \hline
  \end{tabular}
 \end{adjustbox}
\end{table}

\section{Conclusions}

We found that the frame-wise CE training
for transducer models greatly simplifies,
speeds up and improves our transducer training
by methods like chunking.
It also allows us to train a novel transducer model.
We gain interesting insights regarding model behaviour in decoding.
Finally we achieve good results compared to the literature
within much less training time.
Our final transducer model is better than our attention model,
and also generalizes much better on longer sequences.

\section{Acknowledgements}


\begin{spacing}{0.8}
This work has received funding
from the European Research Council (ERC)
under the European Union’s Horizon 2020 research
and innovation programme
(grant agreement No 694537, project ”SEQCLAS”)
and from a Google Focused Award.
The work reflects only the authors’ views and none of
the funding parties is responsible for any use that may be
made of the information it contains.
\end{spacing}

\bibliographystyle{IEEEtran}

\let\OLDthebibliography\thebibliography
\renewcommand\thebibliography[1]{
  \OLDthebibliography{#1}
  \setlength{\parskip}{0pt}
  \setstretch{0.8}  
}

\bibliography{rnnt-paper}

\end{document}